Mechanical and Electrical Oscillations of a Superconductor Coil and its Applications.

Osvaldo F. Schilling

Departamento de Física, Universidade Federal de Santa Catarina, Campus, Trindade, 88040-900, Florianópolis, SC. Brazil.

email: osvaldof@mbox1.ufsc.br      fax: +55-483319946

**In this paper we discuss the phenomenon of resonance that we predict will occur when a superconducting coil is submitted simultaneously to a magnetic field and to an external force. When both the force and the magnetic field are constant in time the coil should oscillate mechanically around an equilibrium position like an ideal harmonic oscillator, and should have induced in it an alternating current of same frequency. We obtain an expression for the resonance frequency, which is directly proportional to the applied magnetic field and depends also on the wire superconducting properties. We discuss also a setup with no built-in magnet, in which the coil is driven by a mechanical oscillator, and thus the force varies cyclically with time. A measurable alternating current may be induced in the device for an amount of magnetic flux as small as one flux quantum, so that the device is an excellent magnetometer.**

One of the most widely exploited properties of superconductors is perfect conductivity, which is applied in the fabrication of wires and cables for the loss-less transport of electrical currents. Although not explicit in the idea of loss-less electronic motion, as we show below this concept carries in it the potential of development of a method for the electromechanical conversion of the absence of energy dissipation. Therefore, the objective of this paper is to discuss theoretically, for the first time ( as far as we know) a phenomenon in which the loss-less electron motion of a superconductor may be directly transformed in loss-less mechanical motion. The object of the discussion is a simple system consisting of a magnet and a rectangular superconducting coil. We demonstrate that the system naturally oscillates and reaches resonance when the superconducting coil is submitted simultaneously to a constant magnetic field and to an external force. The coil will oscillate mechanically around an equilibrium position like an ideal harmonic oscillator, and will have induced in it an alternating current of same frequency. We obtain an expression for the resonance frequency, which is directly proportional to the applied magnetic field and depends also on the wire superconducting properties through London's parameter $\Lambda$. We discuss also a setup with no built-in magnet, in which the coil is driven by a mechanical oscillator, and thus the force varies cyclically with time. There is no current in the absence of field, but an alternating current may be induced in the device for a magnetic flux threading the coil as small as one flux quantum so that the device is an excellent magnetometer.

Let´s consider the experimental setup described in Figure 1. A rectangular superconducting coil of mass $m$ is submitted to a uniform magnetic field $B$ from some magnetic source. The coil is pulled away from the magnet by an external force $F$, which in this initial example is assumed independent of time. According to Faraday's Induction Law the motion of the coil in the presence of $B$ will induce in it a transport supercurrent $i$. Such current will generate a magnetic force with strength $F_m = iaB$ in the upper side ( of length $a$) of the coil in opposition to $F$. The coil will move with speed $v$ described by Newton´s Law:

$$m \, d \, v/dt = F - iaB \qquad (1)$$

The displacement of the coil gives rise to an induced electromotive force $\varepsilon$, given by Faraday's induction law[1]:

$$\varepsilon = \int \boldsymbol{E} \bullet d\boldsymbol{s} = -d\Phi_m/dt - L \, di/d \, t \qquad (2)$$

Here $\boldsymbol{E}$ is the electric field in the coil wire, $\Phi_m$ is the magnetic flux from the external source that penetrates the rectangular area bound by the coil, and $L$ is the self-inductance of the coil. The line integral is calculated around the perimeter of the coil. In (2), $d\Phi_m/dt = -Bav$. The electric field in the superconducting wire of the coil must obey the London equation $\partial/\partial t\,(\Lambda J) = E$. Here $J$ is the current density, and $\Lambda = m_e/(n\,e^2)$ is the London parameter, in which $m_e$ is the electron mass, $n$ is the density of quasiparticles that form pairs and carry the supercurrent, and $e$ is the electronic charge [1]. Let´s assume as a simplification that the London penetration depth of the superconducting currents inside the wire is greater than the radius of the wire, so that $J$ may be approximated by $i/A$, where $A$ stands for the cross-sectional area of the wire. $J$ is constant around the perimeter $p$ of the coil. In this case the line integral of $E$ around the coil is:

$$\varepsilon = (\Lambda p/A)\,di/dt \qquad (3)$$

Inserting (3) into (2) one obtains a relation between $v$ and $di/dt$. Taking the time derivative of (1) and eliminating $di/dt$ from (2) and (3) one obtains:

$$m\,d^2v/dt^2 = -(\,B^2a^2/(L + \Lambda p/A))\ \ v \qquad (4)$$

This means that the coil should perform an oscillating motion under the action of the external and magnetic forces. Assuming zero initial speed and an initial acceleration equal to $F/m$ equation (4) can be solved:

$$v(t) = F/(m\Omega)\ \ sin\,(\,\Omega\,t) \qquad (5)$$

The natural frequency of the oscillations is $\Omega = Ba/(m(L+p\Lambda/A))^{1/2}$. In addition it is possible to combine (1)–(3) to obtain an equation for the current $i(t)$:

$$(Ba/\,\Omega^2)\,d^2i/\,dt^2 = F - Bai \qquad (6)$$

whose solution is

$$i(t) = (\,F/\,(Ba))\,(\,1 - cos(\Omega t)) \qquad (7)$$

for $i(0) = di/dt(0) = 0$. It is important to stress that the mechanical motion predicted can be made perfectly (!) frictionless since in the vertical position it might be entirely independent of the existence of any physical contact between the coil and the magnet. Another important point to be discussed is the role of normal electrons in the transport of the alternating current, since this would result in energy losses. We note that the number of free parameters of the model allows the frequency $\Omega$ to be small, so that the influence of eddy currents due to normal carriers might be entirely neglected. Therefore, it is possible to conclude that the superconducting coil-magnet-force system, for all practical purposes, should conserve its initial energy.

In addition to its importance as a basic-science model of an electromechanical energy-conserving system, the setup may have other more practical applications. We will mention two of them. We notice first that the measurement of the resonance frequency $\Omega$ is a function of the London parameter $\Lambda$, which is directly related to the density $n$ of superconducting carriers. Therefore, it might be possible in principle to obtain such important material property as a function of temperature and/or wire composition ( for instance) by measuring $\Omega$. We should point out, however, that the factor $p\Lambda/A$ is actually quite small compared to usual values of $L$ for wires of ordinary thickness. However, if the actual coil is manufactured with a superconductor metallic film of, e.g., 100 $nm$ thickness and width the parameter $A$ will be small enough to make the $p\Lambda/A$ term of the same order of magnitude as the inductance $L$, and thus the parameter $\Lambda$ might easily be measured.

It is interesting to go further and investigate a second technological applications of this experimental setup. We note that a simple variation may produce a very sensitive magnetometer. Let's consider the case in which $F = F_o \sin(\omega t)$ . We may write a set of equations similar to those for the constant force case, to obtain the differential equation for a forced harmonic oscillator. The solution for the speed $v(t)$ is ( neglecting the $p\Lambda/A$ term for simplicity):

$$v(t) = (\omega F_o /(a^2 B^2/L - m\omega^2)) \cos(\omega t) \qquad (8)$$

which means that the amplitude of the oscillating motion will be

$x_o = F_o /(a^2 B^2/L - m\omega^2)$. One may write a differential equation for the current $i(t)$ and obtain:

$$i(t) = (F_o (aB/L)/(a^2 B^2/L - m\omega^2)) \sin(\omega t) \qquad (9)$$

If $a^2 B^2/L \ll m\omega^2$ $(\Omega \ll \omega)$ it is clear from (9) that the absolute value of the amplitude of the alternating current $i_o$ will be proportional to $B$, that is,

$$i_o = F_o aB/(m\omega^2 L) = x_o aB/L \qquad (10)$$

Let's estimate the magnitude of $i_o$ for some typical values of the parameters, that is, $a = 1$ *mm*, $L = 10^{-9}$ *H*, $B = 0.01$ *T*, $m = 0.01$ *kg* ( the coil might be attached to a base or substrate to increase the mass of the system), $\omega = 10$ *rad/s*. For these parameters the condition $\Omega << \omega$ is satisfied. Taking $x_o = 0.1$ *mm* one obtains from eq. (10) that $i_o = 100\ B$, with the current in Ampère and field in Tesla. A field of $10^{-8}$ *T* ( corresponding to about one flux quantum penetrating the $10^{-7}$ *m²* of coil swept by the oscillations) would produce an alternate current of 1 microAmpere, something easily measurable.

In conclusion, we have predicted theoretically that a superconducting coil will perform an effectively loss-less oscillating motion of frequency $\Omega$ around an equilibrium position when simultaneously submitted to a constant force and a constant magnetic field. Such oscillating motion is accompanied by a supercurrent which has an alternating component of same frequency $\Omega$. We note that the frequency might be made so low that any dissipative effects due to normal electrons might be considered negligible. The measurement of $\Omega$, which depends on the applied magnetic field as well as on several geometrical parameters might be used to obtain the London parameter $\Lambda$, and thus the density of superconducting electrons in the wire. We have also discussed the utilization of the setup as a magnetometer, in which case the coil is driven by a mechanical low-frequency actuator.

The author wishes to thank Prof. Said Salem Sugui Jr. and Prof. Mauro M. Doria for helpful discussions.

Reference.

Figure caption.

Figure 1: A rectangular coil is submitted simultaneously to an external force $F$ and a constant magnetic field $B$ perpendicular to it. As shown in the text the predicted motion is oscillatory, with an alternating current of same frequency being induced in the coil.

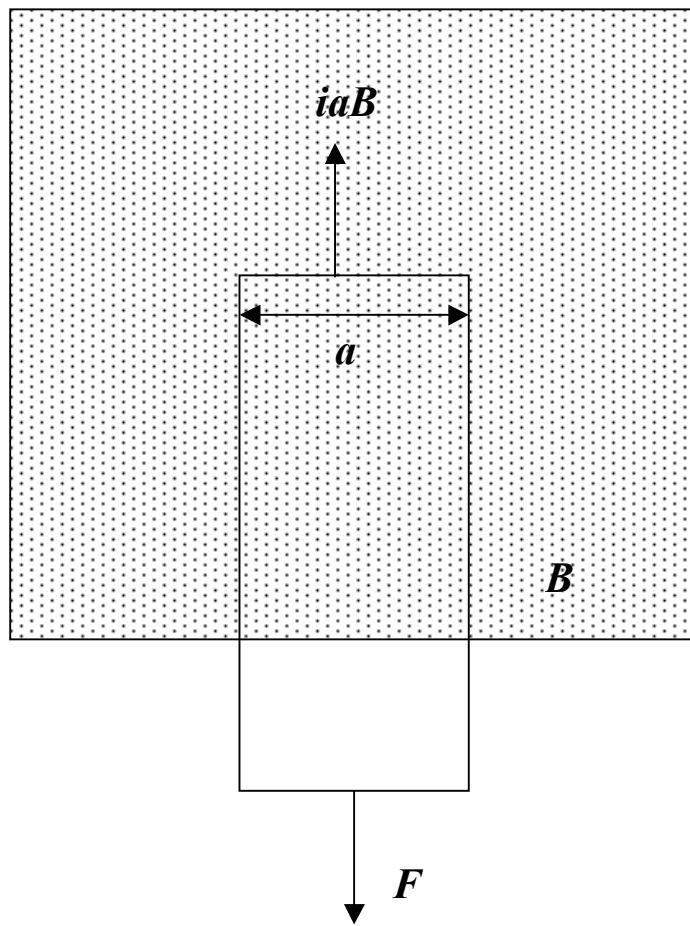

Figure 1
Schilling